\documentclass[aps,prl,preprint,groupedaddress, showpacs]{revtex4-1}
\usepackage{graphicx}
\usepackage{amsmath}
\usepackage{amssymb}
\usepackage{amsthm}

\begin{document}
\thispagestyle{headings}

\title{Probing the dynamics of Anderson localization through spatial mapping}

\author{Ramy G. S. El-Dardiry}
\email[]{dardiry@amolf.nl} \homepage[]{http://www.randomlasers.com}
\affiliation{FOM Institute AMOLF, Science Park 104, 1098 XG
Amsterdam, The Netherlands}

\author{Sanli Faez}
\affiliation{FOM Institute AMOLF, Science Park 104, 1098 XG
Amsterdam, The Netherlands}

\author{Ad Lagendijk}
\affiliation{FOM Institute AMOLF, Science Park 104, 1098 XG
Amsterdam, The Netherlands}

\date{\today}

\begin{abstract}
We study (1+1)D transverse localization of electromagnetic radiation
at microwave frequencies directly by two-dimensional spatial scans.
Since the longitudinal direction can be mapped onto time, our
experiments provide unique snapshots of the build-up of localized
waves. The evolution of the wave functions is compared with
numerical calculations. Dissipation is shown to have no effect on
the occurrence of transverse localization. Oscillations of the wave
functions are observed in space and explained in terms of a beating
between the eigenstates.
\end{abstract}


\maketitle

Recent years witnessed a renaissance in experimental studies on
Anderson localization. This phenomenon, conceived by P. W. Anderson
in 1958 \cite{Anderson1958}, originally described the absence of
diffusion of electrons in random lattices due to interference. Since
Anderson localization is in essence a wave phenomenon, physicists
have successfully extended the scope of localization studies to
electromagnetic waves \cite{Anderson1985, John1987, Genack2000,
Genack2011}, ultrasound \cite{vanTiggelen2008}, and matter waves
\cite{Aspect2009, Inguscio2009, DeMarco2011}.

Similar to other phase transition phenomena, dimensionality plays an
important role. For $d \leq 2$, all states are localized, whereas
for $d = 3$ a phase transition from diffusive to localized behavior
occurs at a critical scattering strength \cite{GangOfFour1979}. In
the special case of transverse localization, formulated by De Raedt
et al. \cite{DeRaedt1989}, one dimension is designed not to be
disordered, whereas disorder is introduced in the other
dimension(s). As a consequence, waves spread out in the
disorder-free dimension, but are confined in the other dimensions.
Waves are always localized in the transverse directions as long as
the transverse system length $L$ is larger than the localization
length $\xi$. Effectively, transverse localization reduces the
number of spatial coordinates in the system: the coordinate along
which the sample is extruded can be seen as the time-axis in the
time-dependent Schr\"{o}dinger equation. Stationary transverse
localization experiments could thus provide a unique insight into
how a localized wave develops over time. Studying and understanding
this intriguing aspect of transverse localization experimentally is
the central topic of this paper.

Pivotal experiments on weakly scattering disordered photonic
lattices \cite{Schwartz2007, Silverberg2008, Segev2011} have
focussed on the observation of localized wave functions after a
certain fixed propagation distance and the effect of nonlinearity on
the transverse localization length. Both theoretical and
experimental studies have revealed interesting dynamical properties
of the periodically kicked quantum rotator which bears close
resemblance to Anderson localization \cite{Prange1982, Raizen1995},
suggesting that studying the dynamics of localization itself is
important. The unique property of transverse localization
experiments that enables us to map one spatial dimension onto time
is ideally suited for this purpose.

\begin{figure}
  \centering
  \includegraphics[width=0.6\textwidth]{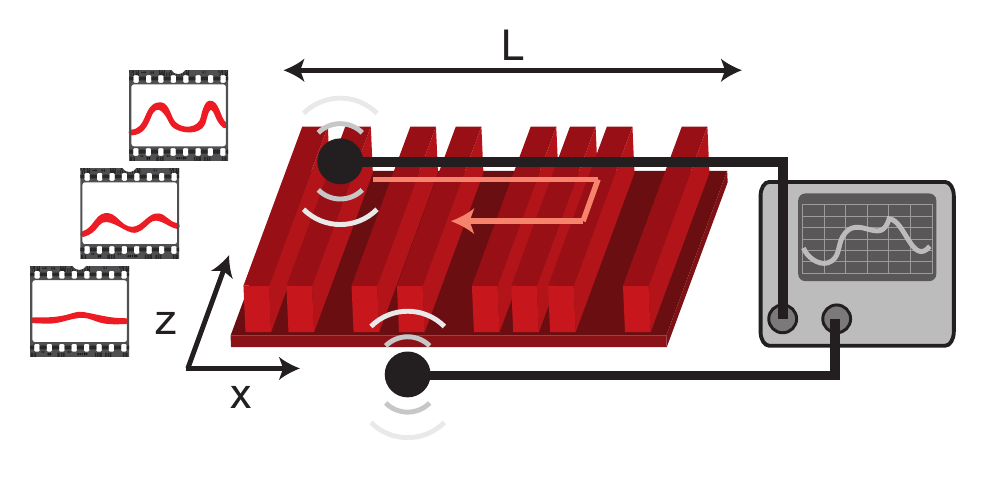}\\
  \caption{Experimental set-up. Nylon bars (red) are placed on top
  of an oxygen free copper plate, the distance between the bars is random in the
  transverse x-direction. The z-direction is disorder free. One of
  the two microwave antennas (black disks) is scanned over the sample and a vector network analyzer is used to measure the transmitted
  spectrum
  between the two antennas. Each scan along the x-axis is
  equivalent to a snapshot in time.
  }\label{gExperiment}
\end{figure}

First, we provide new results on transverse localization by
performing experiments with microwaves in 2D. Instead of measuring
the intensity at the end of our samples, we study the evolution of
the extent of the waves as a function of propagation distance.
Second, we compare our experimental results with numerical solutions
to a 2D Schr\"{o}dinger-like equation from which we deduce
information regarding the effect of dissipation on localization.
Since localization was introduced for classical waves the issue of
absorption has been the subject of immense discussions and various
opinions \cite{Anderson1985, John1985}. Tackling this issue has
shown to be unavoidable in any experiment \cite{Weaver1993,
Beenakker1996, Wiersma1997, Maret1999, Genack2009}. To study both
the dynamics of localized waves and the effect of unavoidable
dissipation, we performed measurements on samples consisting of
scattering bars placed parallel to each other in an open system.
Such samples form an excellent model system in which out-of-plane
scattering plays the role of dissipation, whereas propagation along
the bars represents the dynamics. Interesting non-stationary
behavior of single localized wave functions is observed, which we
analyze by decomposing these functions into the system's eigenstates
semi-analytically.


\textit{Experimental methods - } Figure \ref{gExperiment} shows a
sketch of the experimental apparatus. Samples were fabricated by
placing nylon bars (3 mm $\times$ 10 mm $\times$ 1000 mm) on top of
an oxygen free copper plate (500 mm $\times$ 1000 mm). These nylon
bars ($n=1.73$ \cite{NylonHP}) are the scatterers in our system.
Disorder was introduced into the system by varying the spacing
between the nylon bars. The spacings were chosen randomly from a
Poissonian distribution with a mean of 10 mm. Introducing Poissonian
disorder makes sure that the presence of stop band effects is
negligible \cite{Segev2011b}. In addition, ordered samples were
prepared with a lattice spacing of 20 mm in which clear stop bands
were observed and calculated around 7 and 13 GHz. Styrofoam spacers
ensured parallel alignment of the nylon bars. We studied the sample
by measuring the microwave transmission spectrum with varying
bandwidths around 10 GHz using a vector network analyzer (Rhode and
Schwartz ZVA 67). The detection antenna was scanned over the sample
by using a stepper motor (Newport ESP 301) and a home-built scanning
stage.

\begin{figure}
  \centering
  \includegraphics[width=0.65\textwidth]{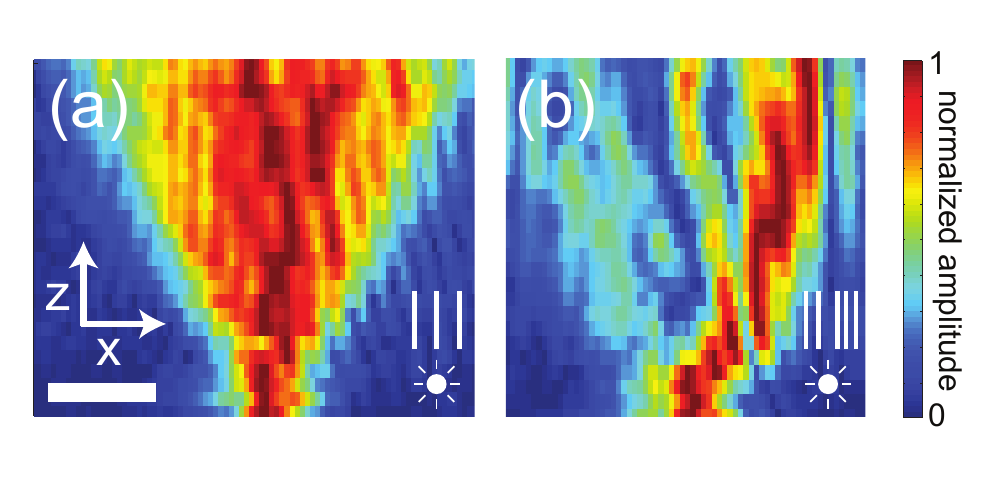}\\
  \caption{Experimentally determined false color images of the amplitude
  distribution for (a) an ordered and (b) a disordered sample at 9.2 GHz . Every row is normalized
  independently. The scale bar denotes 100 mm.
  }\label{gordvsdis}
\end{figure}

\textit{Results - } The propagation of waves within an ordered and a
disordered sample is shown in Fig. \ref{gordvsdis}. The excitation
frequency was set at 9.2 GHz, that is outside any of the stop gaps
of the ordered sample. The data was normalized for every row in the
\textit{xz}-plane to enhance the visibility of the wave function far
away from the source. In the ordered sample, Fig.
\ref{gordvsdis}(a), waves spread out ballistically as a function of
propagation distance. However, for the disordered sample, Fig.
\ref{gordvsdis}(b), the wave propagation is strikingly different:
the wave initially spreads out, but at a certain stage stays
confined to a bounded region. These type of two-dimensional spatial
scans provide us with exceptional data for analyzing transverse
localization in unprecedented detail.

In order to quantify the transverse confinement of wave intensity as
function of propagation distance, we calculate the inverse
participation length (IPL) \cite{Mirlin1993}. 
The IPL for a one-dimensional intensity distribution $I(x)$ is
defined as
\begin{align}\label{eqIPR}
P(z)\equiv\frac{\int I^2(x,z){\rm d}x}{\left(\int I(x,z) {\rm
d}x\right)^2}
\end{align}
and has a unit of inverse length. The IPL is inversely proportional
to the spread of the wave function: a homogeneously extended wave
spread out over the entire sample length $L$ leads to an IPL of
$1/L$. To obtain a reliable value for the spread of wave functions,
the ensemble averaged intensity profiles were determined by
averaging over 20 realizations of disorder.

\begin{figure}
  \centering
  \includegraphics[width=0.65\textwidth]{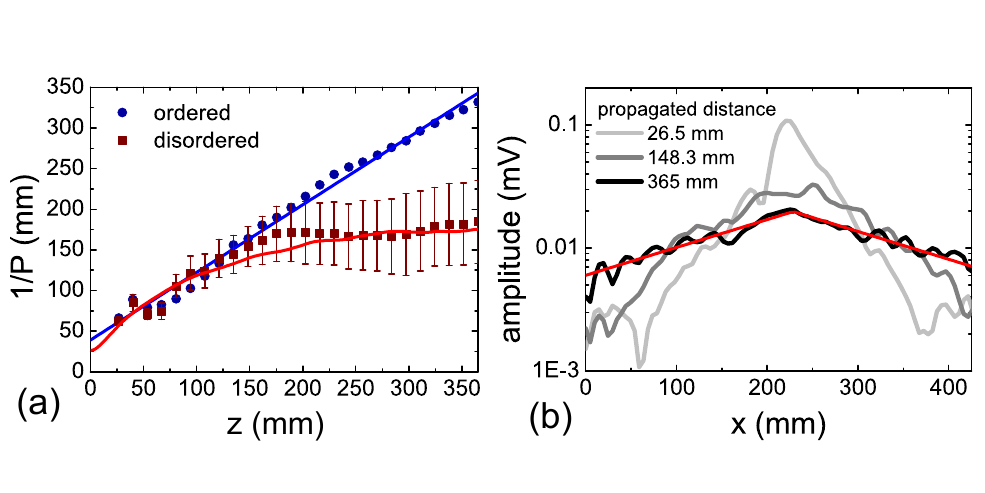}\\
  \caption{(a) Participation ratio versus propagation distance at
  9.2
  GHz for an ordered sample (blue) and a disordered
  ensemble (red). Red line: calculation for an ensemble of 100 disordered samples. Blue line: linear fit.
  (b) Transverse ensemble averaged amplitude distribution for different propagation
  distances. Red line: exponential fit.
  }\label{gpartratio}
\end{figure}

Figure \ref{gpartratio}(a) shows how the inverse of the IPL develops
with increasing propagation distance for both the ordered sample and
the ensemble of disordered samples at 9.2 GHz. In agreement with the
qualitative picture we obtained from Fig. \ref{gordvsdis}, we see
that the extent of the wave function given by the inverse of the IPL
increases linearly for the ordered sample. For the disordered
ensemble on the other hand the IPL flattens off after a certain
propagation distance. This settling of the IPL to a finite value
constitutes a direct experimental observation of the spatial
evolution of transversely localized waves.

Besides a different evolution of the waves' extent, the eventual
spatial shape of the ensemble averaged wave function distinguishes
localized from extended waves as well. In contrast to Gaussian
shaped extended wave functions, localized wave functions obtain
exponential tails. Figure \ref{gpartratio}(b) shows the ensemble
averaged wave function profile for three different propagation
distances on a semi-logarithmic scale. Close to the excitation
source, the ensemble averaged wave function is strongly peaked in
the transverse dimension. For longer propagation distances, the
intensity in the wings of the wave function increases and the peak
becomes less pronounced. The ensemble averaged wave function
quenches once the localization length is reached and its shape is
well described by an exponential. From an exponential fit to the
data we find a localization length of 192 $\pm$ 6 mm.

\begin{figure}
  \centering
  \includegraphics[width=0.65\textwidth]{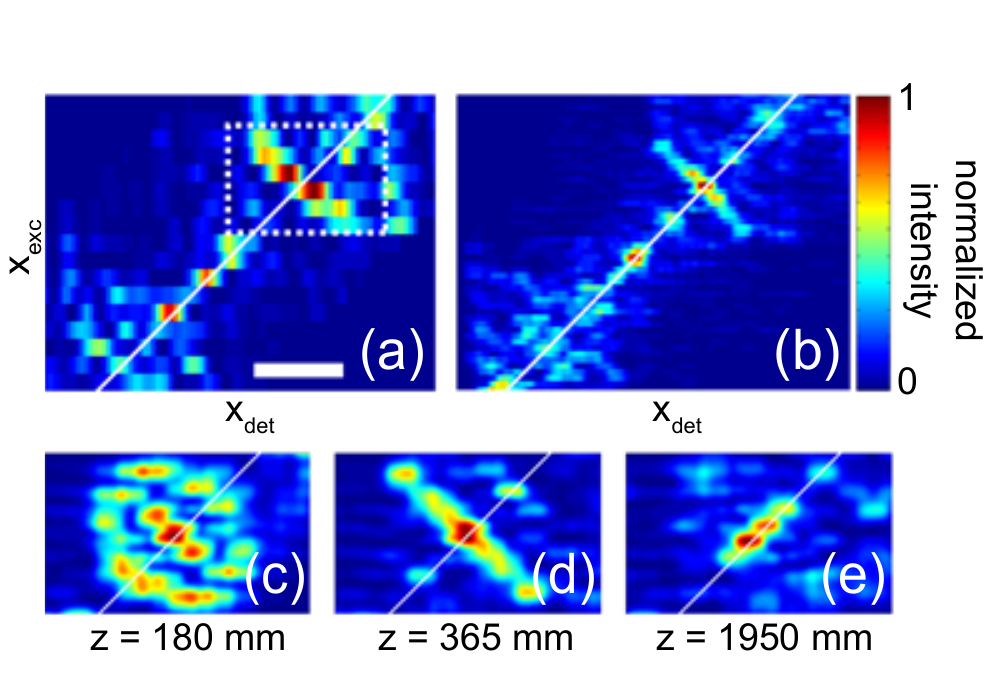}\\
  \caption{(a) Experimental and (b) numerically calculated false color plots of the wave function
  intensity in transverse direction after 365 mm of propagation
  along the z-direction
  for different positions of the excitation antenna at 9.2 GHz. The white lines
  indicate the position of the excitation antenna. The dashed box
  marks an anti-diagonal wave profile.
  Scale bar denotes 100 mm. (c-e) show calculations
  using mode decomposition for the area marked with the dashed box
  in (a) for 180, 365, and 1950 mm of propagation respectively.
  Beating of eigenmodes can result in (c) circular, (d)
  anti-diagonal, or (e) diagonal patterns.
  }\label{gexcitation}
\end{figure}

After having studied these ensemble averaged properties of our
system, we now aim to understand the propagation of waves for single
realizations of disorder. In Fig. \ref{gexcitation}(a), we plot the
spatial profile for 17 different excitation positions in one sample
after 365 mm of propagation. Based on Fig. \ref{gpartratio}, this
distance ensures we are truly looking at localized wave functions.
The individually measured spatial profiles are strongly dependent on
the position of the excitation antenna. To a large extent the
detected radiation follows the position of the excitation antenna as
indicated by the white diagonal. Naively, one might expect for a
localizing sample clearly isolated regions of higher intensity that
are independent on the position of excitation. Such patterns would
appear as vertical stripes in Fig. \ref{gexcitation}(a). However,
much to our surprise, the spatial patterns of these isolated regions
along the transverse dimension $x$ are dependent on the excitation
position. In fact, for the measurement shown in Fig.
\ref{gexcitation}, some patterns appear to be anti-diagonal.
Furthermore, the excitation map displays a high degree of symmetry
along the diagonal.

\textit{Model - } In order to build a basis for understanding the
ensemble averaged data and the remarkable excitation dependence of
localized wave functions in single realizations of disorder, the
system is analyzed numerically and semi-analytically. In the
paraxial limit transverse localization is described by an equation
which closely resembles the time-dependent Schr\"{o}dinger equation
\cite{DeRaedt1989} where $z$ plays the role of time
\begin{align}\label{eqTransverseEq}
i\frac{\partial \psi}{\partial z} = \frac{1}{2kn_0}H\psi,
\end{align}
with $\psi$ the wave field and $k$ the vacuum wave number. The
effective index of refraction is given by $n_0^2 \equiv L^{-1}
\int_L n^2(x)dx$. The Hamiltonian is defined as
\begin{align}\label{eqHamiltonian}
H\equiv\frac{\partial^2}{\partial x^2} + k^2[n^2(x)-n_0^2].
\end{align}
A term $-i\alpha$ can be added to $H$ creating an effective
Hamiltonian that also describes losses due to dissipation or
out-of-plane scattering. Partial differential Eq.
(\ref{eqTransverseEq}) is rewritten as a set of ordinary
differential equations in $z$ by using the method of lines
\cite{Schiesser}. After separating the real and imaginary part of
$\psi$, we use MatLab to solve the equation numerically by means of
a Runge-Kutta algorithm. The $xz$-plane is discretized in
600$\times$200 steps. The initial wave at $z=0$ is modeled as a
Gaussian with a width of 1.15 cm given by the aperture of the
excitation antenna. To compare the numerical calculation with
experiment, we convolved the intensity of the calculated wave
function with the aperture of the detection antenna.

\begin{figure}
  \centering
  \includegraphics[width=0.65\textwidth]{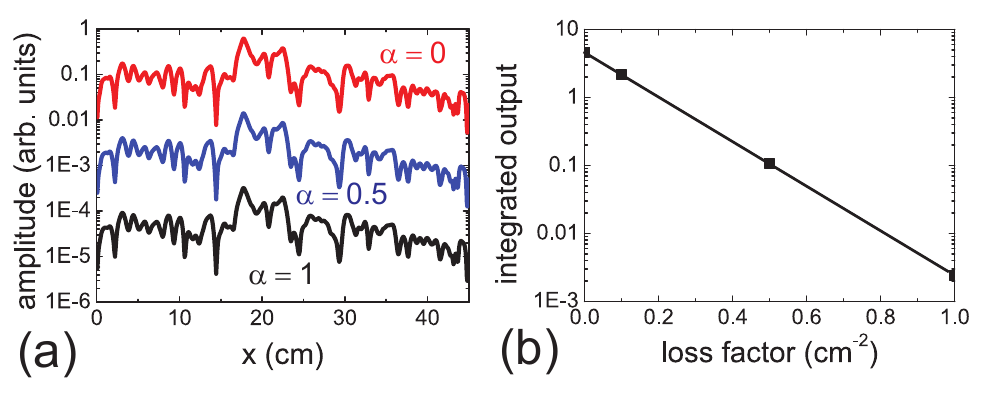}\\
  \caption{(a) Mode amplitude profile versus transverse distance for
  different values of the loss coefficient $\alpha$ after 350 mm of propagation. Red: $\alpha = 0$ cm$^{-2}$.
  Blue: $\alpha = 0.5$ cm$^{-2}$. Black: $\alpha = 1$ cm$^{-2}$.
  (b) Integrated output versus loss coefficient on a semilog scale.
  }\label{gabsorption}
\end{figure}

Our ``open" experimental configuration requires that we first
analyze the role of dissipation on transverse localization. The
amplitude profile is calculated after 350 mm of propagation for a
single realization of disorder for various values of $\alpha$ as
shown in Fig. \ref{gabsorption}(a). For higher values of the
absorption coefficient, the amplitude decreases, but the wave
function shape does not alter. We conclude from these curves that
absorption merely scales the wave intensity thereby supporting our
experimental approach of studying transverse localization in the
possible presence of out of plane losses. In Fig.
\ref{gabsorption}(b), the integrated output is plotted versus the
absorption coefficient. The output intensity clearly attenuates
exponentially, confirming that dissipation only introduces an
exponential scaling \cite{Weaver1993}. In Fig. \ref{gpartratio}(a),
the mean of the participation length for an ensemble of 100
realizations of disorder is shown. This theoretical value for the
waves extent falls within the standard deviation of the
experimentally determined values.

Motivated by the experimentally observed and unforeseen excitation
dependence of the wave functions, we also calculated the
excitation-detection patterns. In Fig. \ref{gexcitation}(b), it is
shown that the position and shape of these patterns correspond with
the measurements. The anti-diagonal shapes are also clearly present
in our numerical calculation, indicating that it is not caused by
spurious effects such as mode perturbation due to the proximity of
the receiver antenna.

\begin{figure}
  \centering
  \includegraphics[width=0.65\textwidth]{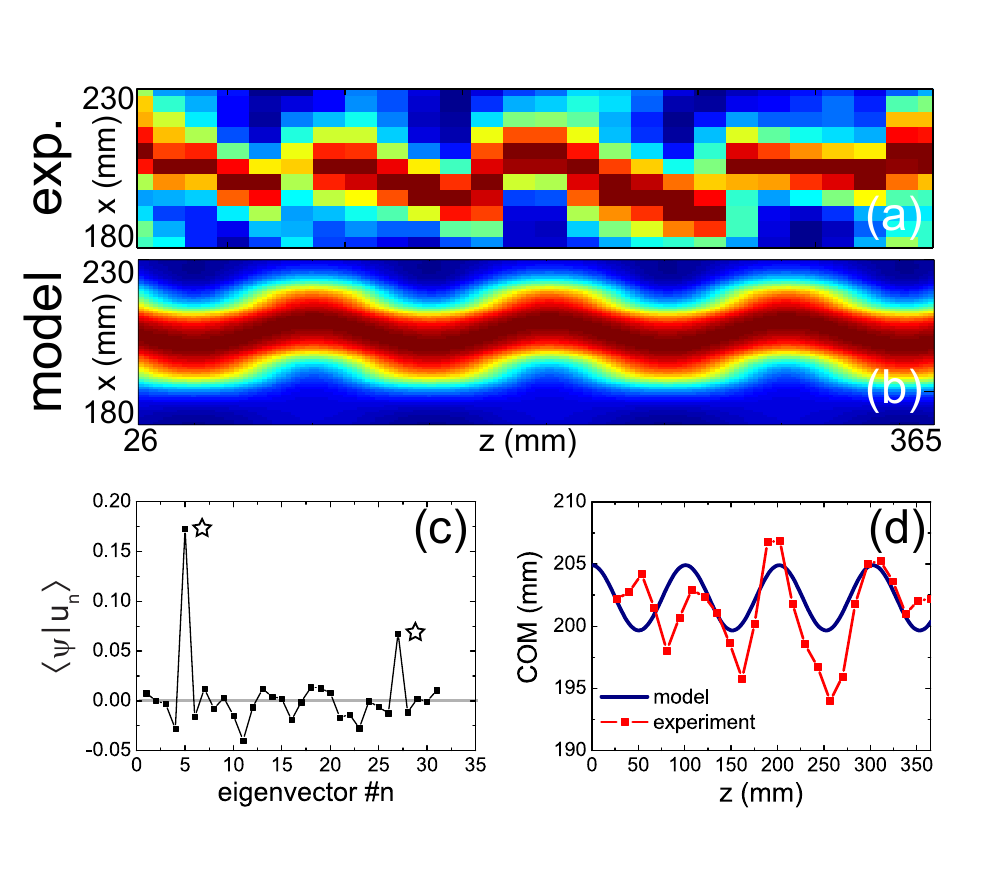}\\
  \caption{(a) Experimental and (b) calculated oscillations in the
  intensity profile for one sample excited at 10.2 GHz. Every column is normalized independently. (c) Expansion coefficients for the different
  eigenstates. Only two eigenstates, indicated by the stars, contribute significantly. (d) Center of mass of the intensity versus propagation direction for
  both the calculated and experimental data shown in (a) and (b).
  }\label{goscillations}
\end{figure}

An alternative analysis of the system in terms of eigenstates rather
than the previous ``brute force" numerical calculation has the
advantage of reducing the problem's complexity. When $\alpha = 0$,
the solutions to Eq. (\ref{eqTransverseEq}) can be written as a
linear combination of the Hamiltonian's eigenstates $u_n(x)$:
$\psi(x, z) = \sum_n c_n u_n(x) {\rm exp}(-i\lambda_n z)$, where
$\lambda_n$ is the eigenvalue belonging to eigenstate $u_n$ and
$c_n$ is the $n$'th expansion coefficient given by $c_n =
\langle\psi|u_n\rangle$. The eigenstates and eigenvalues are
calculated by diagonalizing a 598 $\times$ 598 matrix. The diagonal
of the matrix contains the potential $k^2[n^2(x)-n_0^2]$ and the
derivative in $x$ is approximated by using central differences
creating a tridiagonal matrix when assuming absorbing boundary
conditions.

In principle, diagonalizing a $N\times N$ matrix results in $N$
eigenvalues and eigenvectors. However, most of these eigenvectors
contain too high spatial frequencies, $k_n>k$, that are not
excitable in our system. As a result we end up with only 30
eigenstates that obey the relation $k_n\leq k$. This small number of
modes in the system can lead to observable beatings of the system's
eigenstates in our two-dimensional spatial scans. Figure
\ref{goscillations}(a) shows a clear example of such oscillatory
behavior of the wave function in experiment. A decomposition into
the system eigenstates for this particular sample reveals just two
eigenstates contribute significantly to the wave function as shown
in Fig. \ref{goscillations}(c). Using only these two eigenstates and
their corresponding eigenvalues, we calculated the $z$-development
of the wave function in Fig. \ref{goscillations}(b) and compared
them directly with experiment in Fig. \ref{goscillations}(d). The
calculated oscillations are quantitatively similar to those observed
in experiment. In general, the number of significantly contributing
eigenvectors is often higher than two, which makes the beating less
visible. Yet, the anti-diagonal patterns shown in Fig.
\ref{gexcitation}(a) and (b) are another observable consequence of
the beating between the system's eigenstates. Depending on the
accumulated phase during propagation, these anti-diagonal
excitation-detection patterns can in fact become circular or
diagonal as shown in Fig. \ref{gexcitation}(c-e). The patterns are
to a large extent point-symmetric which originates from a flip in
sign of the expansion coefficients when the excitation antenna
crosses the central position of the beating oscillation.

 \textit{Conclusion and Discussion - } We have
measured how electromagnetic wave functions develop over time in
localizing samples  by carrying out a (1+1)D transverse localization
experiment. Because of the limited number of modes in our system,
the excitation can be described as a superposition of a few of the
system's eigenstates. The different eigenvalues of these eigenstates
lead to observable beatings in wave functions.

We used out-of-plane scattering as an experimental analog of energy
dissipation. The extent of the wave profiles is in quantitative
agreement with calculations from a numerical solution to a
Schr\"{o}dinger type of equation. By introducing dissipation into
this model, we deduce that dissipation is of no influence to the
occurrence of transverse localization except for an exponential
attenuation.


Since the transverse localization scheme allows for measuring
snapshots of wave functions in time, it is a very convenient tool
for studying the effect of different forms of disorder on wave
propagation as put forward by recent work on photonic quasicrystals
\cite{Silverberg2009, Segev2011}. Our work on transverse
localization and dissipation suggests that transverse localization
can also be an excellent platform for studying the influence of
perturbations and partial incoherence on localization
\cite{Buljan2011}.

\begin{acknowledgments}
We thank Bergin Gjonaj and Paolo Scalia for stimulating discussions
and Kobus Kuipers for carefully reading the manuscript. This work is
part of the research program of the ``Stichting voor Fundamenteel
Onderzoek der Materie (FOM)", which is financially supported by the
``Nederlandse Organisatie voor Wetenschappelijk Onderzoek (NWO)".
\end{acknowledgments}

\bibliography{references}

\end{document}